\documentclass[aps,pra,reprint,superscriptaddress,showpacs]{revtex4-1}

\usepackage{watermark}
\usepackage{graphicx}
\usepackage{amsmath,bm}
\usepackage{amsfonts}

\usepackage{hyperref}
\hypersetup{
    bookmarks=true,         
    unicode=false,          
    pdftoolbar=true,        
    pdfmenubar=true,        
    pdffitwindow=false,     
    pdfstartview={FitH},    
    pdftitle={Coherent and stochastic contributions},   
    pdfauthor={G. F. Gribakin},     
    pdfsubject={Compound resonances and doorways},   
    pdfcreator={G. F. Gribakin},   
    pdfproducer={Producer}, 
    pdfkeywords={keywords}, 
    pdfnewwindow=true,      
    colorlinks=true,       
    linkcolor=blue,          
    citecolor=blue,        
    filecolor=magenta,      
    urlcolor=blue           
}

\newcommand{\fref}[1]{Fig.~\ref{#1}}

\newcommand{\Eref}[1]{Eq.~(\ref{#1})}

\newcommand{\rtw}{\longrightarrow}
\newcommand{\Gspr}{\Gamma_\mathrm{spr}}
\newcommand{\eps}{\varepsilon}
\newcommand{\st}{\sigma _{\rm tot}}

\begin{document}

\title{Coherent and stochastic contributions of compound resonances in atomic
processes: Electron recombination, photoionization and scattering}

\author{V. V. Flambaum}
\affiliation{School of Physics, University of New South Wales,
Sydney 2052, Australia}
\affiliation{New Zealand Institute for Advanced Study, Massey University
Auckland, 0745 Auckland, New Zealand}
\author{M. G. Kozlov}
\affiliation{School of Physics, University of New South Wales,
Sydney 2052, Australia}
\affiliation{Petersburg Nuclear Physics Institute, Gatchina 188300,
Russia}
\affiliation{St Petersburg Electrotechnical University ``LETI'', Prof
Popov St 5, St Petersburg 197376, Russia}
\author{G. F. Gribakin}
\affiliation{School of Mathematics and Physics,
Queen's University Belfast, Belfast BT7 1NN, Northern Ireland, United Kingdom}

\date{\today}

\begin{abstract}
In open-shell atoms and ions, processes such as photoionization,
combination (Raman) scattering, electron scattering and recombination,
are often mediated by many-electron compound resonances. We show that their
interference (neglected in the independent-resonance approximation) leads to
a coherent contribution, which determines the energy-averaged total cross
sections of electron- and photon-induced reactions obtained using the optical
theorem. In contrast, the partial cross sections (e.g., electron
recombination, or photon Raman scattering) are dominated by the stochastic
contributions. Thus, the optical theorem provides a link between the
stochastic and coherent contributions of the compound resonances. Similar
conclusions are valid for reactions via compound states in molecules and nuclei.
\end{abstract}

\pacs{
31.10.+z, 
34.10.+x, 
34.80.Lx, 
05.45.Mt
}

\maketitle

\section{Introduction}

The aim of this paper is to examine the interplay between simple, singly or
doubly excited ``doorway'' states, and multiply excited chaotic eigenstates
(compound resonances) in processes, such as photon and electron scattering,
photoionization, electron recombination, etc., involving complex atomic or
molecular systems.
In particular, we
identify the coherent and incoherent contributions of the compound resonances,
and show how these are related to the total and partial cross sections of
various reactions. We outline a method for the calculation of probabilities
of these reactions, which involves summations over the doorway states, rather
than the eigenstates.

\subsection{Many-body quantum chaos} 

Consider a finite quantum system with many degrees of freedom, such as a
many-electron atom or ion, a polyatomic molecule, or a heavy nucleus. In the
zeroth-order approximation, the states of such system can be constructed from
some single-particle states. For atoms these will be the electron orbitals
obtained in some mean-field potential, e.g., using the Hartree-Fock method.
For molecular vibrations, the zeroth-order states are normal-mode vibrations,
which are determined by the quadratic expansion of the ground-state electronic
energy near the equilibrium positions of the nuclei.

In general, this description works well for the ground state of the system, and
in many cases, it also provides a correct picture of low-lying excitations.
Thus, the ground states of most atoms and ions are characterized by their
electronic \textit{configuration}. The ground state of the molecular
vibrational Hamiltonian is simply a product of the zero-point motion states of
all the normal modes. Low-energy excitations then correspond to promotions
of one of the electrons into an excited-state orbital, or adding a vibrational
quantum to one of the normal-mode harmonic oscillators.

Of course, the exact energy of the atomic excitation will be affected
by the \textit{residual} two-body Coulomb interaction between the electrons.
Such correction can be relatively small in atoms or ions with a simple
ground-state configuration (e.g., in alkali-like systems with one active
electron above a closed-shell core). At the same time, in systems with
several valence electrons, and in particular, with open-shell ground-state
configurations, the single-particle picture does not hold well at all. A state
in which one of the electrons is promoted to a higher-lying orbital will be
\textit{mixed} with other excited states, in which two or three
electrons have changed their places. Such effects are usually described as
configuration mixing. Finding the eigenstates of the system then requires
constructing a basis of many-electron states of the relevant electronic
configurations, and diagonalizing the Hamiltonian of the residual interaction
in this basis. Similarly, accurate vibrational energies can be found by
including anharmonic, e.g., cubic and quartic, terms in the vibrational
Hamiltonian and diagonalizing its matrix, constructed using the zeroth-order
(harmonic) basis states.

A practical limitation to this approach is set by the maximum size of a
matrix that can be diagonalized efficiently on a computer. The Hamiltonian
matrix sizes become very large in atomic systems with open $d$ and $f$
shells, due to a large number of active electrons, or in polyatomic molecules
with many vibrational degrees of freedom. Such systems are characterized by
large densities of the energy spectra, which promotes strong mixing of
the zeroth-order basis states. As a result, each of the eigenstates becomes
a superposition of a large number of basis states, with the expansion
coefficients behaving like random variables. Further, these eigenstates often
cannot be assigned any meaningful quantum numbers, except the exact ones, such
as the total angular momentum or parity. Even when the Hamiltonian matrix sizes
are manageable, ``exact'' calculations of the spectra
and processes in such systems are virtually impossible due to the extreme
sensitivity of the eigenvalues to small perturbations, e.g., the effect of
states omitted from the basis, or higher-order corrections to the perturbation.

\begin{figure*}
\centering
\begin{picture}(0,0)
\put(3,164){{$\mbox{\large (a)}$}}
\put(256,164){{$\mbox{\large (b)}$}}
\end{picture}
\includegraphics[width=6.7cm]{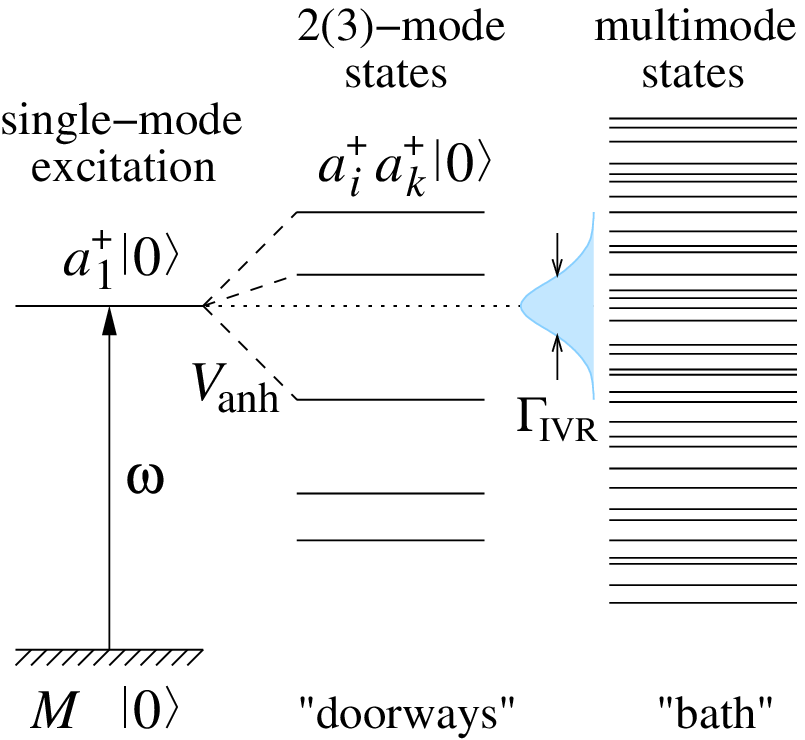}
\hspace{2cm}
\includegraphics[width=7cm]{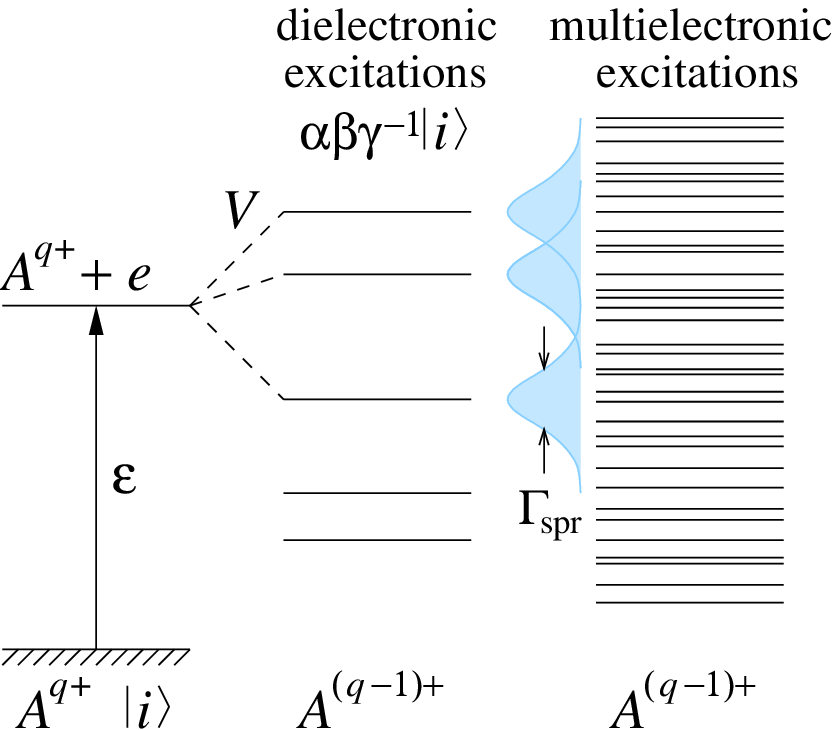}
\caption{Interplay between simple, single-particle degrees of freedom and
multiply excited states in the infrared photoabsorption by a polyatomic
molecule (a) and electron recombination with a many-electron open-shell
ion (b) (see text for details).}
\label{fig:ivr_rec}
\end{figure*}

This behaviour of quantum systems is termed \textit{quantum chaos}. Besides
the Gaussian statistics of the eigenstate components, there is also a specific
correlation between the energy eigenvalues, which is characteristic of
the spectra of random matrices \cite{Guhr1998189}. Well-known examples of
quantum-chaotic systems are excited heavy nuclei (e.g., those formed by neutron
capture) \cite{Zel96,RevModPhys.81.539}, and heavy atoms and ions with open $f$
shells, such as Ce or Au$^{24+}$ \cite{FGGK94,GGF99,GS03}.
Another example is given by the vibrational
motion of polyatomic molecules where anharmonic mixing between normal modes
leads to intramolecular vibrational redistribution (IVR)
\cite{LSP94,NF96,GB98,MMR12}.
Chaotic resonances have also been found recently in ultracold collisions of
erbium atoms \cite{Er} (a manifestation of chaotic states in the excited Er$_2$
molecule).

In each of these examples the quantum-chaotic behaviour of the system leads to
important observable effects beyond the energy-level statistics. Narrowly
spaced neutron resonances in heavy nuclei provide strong enhancements of parity
nonconservation due to the weak interaction \cite{SF1982,FG95}. Electron
capture in chaotic multielectronic resonances in open-shell ions results in
recombination
rates $10^2$--$10^3$ times greater than the single-particle radiative
recombination rate
\cite{GGF99,FGGH02,PhysRevA.86.022714,PhysRevA.88.062713},
as seen in many experiments
\cite{UMSL96,PhysRevA.57.4365,HUSF98,PhysRevA.83.012711,PhysRevA.85.052716,PhysRevA.90.032715}.
Similar states feature in photoionization and photoemission in many ions,
producing a complex
interplay of broad and narrow resonances \cite{MST98,SCD99,LKM05,AGG06,PhysRevA.79.042507,HEP09,SRB14}.
IVR is an essential step in most chemical reactions. It also plays a key role
in electron attachment and positron annihilation in polyatomic
molecules \cite{PhysRevA.77.042724,RevModPhys.82.2557}.

\subsection{Doorway states}

While the exact calculation of many-body chaotic eigenstates is impossible,
their nature allows one to develop a statistical theory to calculate
the mean-squared values of matrix elements and amplitudes involving
such states \cite{Flambaum:93,FlVo93,FGGK94,FG95,FGG98,FGGP98,FGGP99,FG2000}.
In this way one can predict observables averaged over a small energy interval
containing many such states (which is often sufficient since the individual
states cannot be resolved experimentally).

Of particular importance in this approach are \textit{doorway} states.
A doorway is a state which is coupled in the lowest order to the initial state.
For example, in electron-ion recombination, doorways are dielectronic states
(i.e., 1-hole-2-particle excitations of the combined ion). In molecular
infrared photoabsorption, doorways are single-mode excitations. In considering
the IVR process of a single-mode excitations, the doorways are two- and
three-quantum vibrational excitations coupled to the single-mode excitation
in the lowest order.


To illustrate these examples, \fref{fig:ivr_rec} (a) shows schematically the
absorption of an infrared photon of energy $\omega $ by a polyatomic molecule,
followed by IVR.
Figure \ref{fig:ivr_rec} (b) describes the recombination of an electron with
energy $\eps $ with a multicharged positive ion $A^{q+}$.

In the description of both processes, we adopt a \textit{temporal} picture of
the dynamics, as if probed by a short initial pulse. This picture is observed
directly in the pump-probe studies of molecular IVR \cite{Yamada04}. In
contrast, the electron-ion recombination usually deals
with incident electrons of definite energy. (In spite of the high energy
resolution achieved in experiments with electron coolers in ion storage rings
\cite{PhysRevA.57.4365,PhysRevA.83.012711,Glans1999,PhysRevLett.86.5027}, the
measurements for complex targets are incapable of resolving individual chaotic
resonances \footnote{It is interesting to compare the resolution achieved
in the state-of-the-art electron-ion recombination measurements with that
possible in the studies of molecular spectra. In the former case, the
resolution is limited by the temperature of the electron beam, which can be
as low as 1--10 meV \cite{Muller1999,Glans1999,PhysRevLett.86.5027}, which
gives the relative
precision of $10^{-5}$, when compared with the excitation energy of the ion
(e.g., Au$^{24+}$ or W$^{19+}$) formed in the recombination event. In contrast,
the frequency spread of $10^{-4}$~cm$^{-1}$ achieved in high-resolution
molecular spectroscopy \cite{McIN90}, is about $10^{-8}$ of the vibrational
energy probed, so the experiment is capable of detecting individual ``bath''
states.}.)

\subsubsection{Vibrational excitation of molecules}

In the process shown in \fref{fig:ivr_rec} (a), the energy of the photon is
tuned to the frequency of the normal mode 1 (e.g., a CH or OH stretch mode,
with $\omega \sim 3000$~cm$^{-1}$). In the first step the photon
excites a single-quantum vibration of this mode. Lowest-order (cubic and
quartic) anharmonic
couplings $V_{\rm anh}$ perturbatively couple this initial state to some
two- or three-mode vibrational excitations. These off-resonance states act as
``doorways'' which mediate the \textit{spreading} of the vibrational energy
into more complex multi-mode vibrational states, whose density is much higher
than that of the modes or doorways. If $V_{\rm anh}$ is sufficiently strong
(and suitable doorway states are available), the excitation ultimately spreads
into the ``bath'' of closely-spaced states \cite{StuMa93}.

This is the essence of the IVR
process. Its time scale $\tau \sim \hbar /\Gamma _{\rm IVR}$, is related to the
energy width of the initial single-mode state with respect to its decay
towards the bath states. High-resolution molecular spectroscopy in fact allows 
one to observe these states as clumps of narrowly spaced absorption lines
within $\Gamma _{\rm IVR}$ energy interval of the vibrational fundamental
\cite{McIN90}. The number of such lines is
$N\sim \Gamma _{\rm IVR}\rho _v$, where $\rho _v$ is the total density of the
vibrational spectrum for a given symmetry, at this energy.

\subsubsection{Electron-ion recombination}

Turning to the second process [\fref{fig:ivr_rec} (b)], the ground state of the
target ion $A^{q+}$ is usually a simple state described by a single dominant
electronic configuration. The Coulomb interaction $V$ between the incident and
target electrons couples the initial state ($e^-+A^{q+}$) to the doubly excited
states of the compound ion $A^{(q-1)+}$. In such states two electrons occupy
some excited-state orbitals ($\alpha $ and $\beta $), leaving a hole in one
of the target ground-state orbitals ($\gamma $). For simple targets,
photoemission from the doubly excited state completes the
\textit{dielectronic recombination} process \cite{MB42}.

For open-shell targets
such as Au$^{25+}$ (with $4f^8$ outer orbital ground-state configuration),
the dielectronic resonances are embedded in a dense spectrum of multiply
excited states and are strongly mixed with them \cite{GGF99,GS03}. In the
temporal picture this mixing describes a rapid decay of the
dielectronic excited states into chaotic ``compound states'' (a term which
originated in nuclear physics, sometimes called Feshbach resonances \cite{Feshbach62}). Its time constant $\tau $
is determined by the so-called \textit{spreading width} $\Gspr$, as
$\tau =\hbar/\Gspr $. (It plays the same role as $\Gamma _{\rm IVR}$ in the
first example, but on a completely different scale, e.g., $\Gspr \sim 10$~eV
in Au$^{24+}$.) In the energy eigenstate picture,
each of the dielectronic states appears as a component in many chaotic compound
states, contributing significantly to $N\sim \Gspr /D$ of them ($D$ being the
small level spacing between the compound states).

As a result of this spreading, the weight of every doorway in a given
compound state is $\sim 1/N$, and the probability for the compound states to
autoionize (i.e., re-emit the electron) is greatly reduced ($\propto N^{-1}$).
On the other hand, their lifetimes with respect to emitting a photon are
similar to those of the dielectronic (and singly excited) states, since any
excited electron in the compound state can radiate. The electron
``trapping'' in the chaotic compound states thus leads to strongly increased
recombination rates \cite{GGF99,FGGH02}.
Specific examples of doorway states for electron recombination with Au$^{25+}$
can be found in Table I of Ref.~\cite{FGGH02}, while mixing of doorways with
chaotic states was explored in Ref. \cite{GS03}.

The energy spacing between the compound states can be very small (see, e.g.,
the estimates for Au$^{24+}$ in Refs. \cite{GGF99,FGGH02}), beyond the
best resolution available in the recombination experiments \cite{Note1}. This
does not mean, however, that the recombination cross section is completely
structureless. The dielectronic doorway states can produce broad maxima with
widths $\sim \Gspr $ in the energy dependence of the cross section. This is
similar to the way in which the frequencies and strengths of vibrational
fundamentals
determine the overall infrared absorption spectrum of a polyatomic molecule.
Here the normal modes excited by the photon play the role of doorways for the
IVR which follows molecular photoabsorption.

In what follows we consider a variety of processes initiated by a photon or
electron impact on a complex atomic or molecular system. We aim to determine
the roles played by the simple doorway states and chaotic, compound states in
each case. Although most of the expressions and conclusions are quite general,
we will use the language of atoms (or ions) and atomic processes, with the
many degrees of freedom and complexity (chaos) arising from the large numbers
of active electrons and available orbitals.

\section{Theory}

\subsection{Compound states.}

In isolated quantum many-body systems chaos emerges due to a rapid, exponential
growth of the level density with energy. This growth is caused by the increase
in the number of active particles promoted into unoccupied orbitals,
following the increase in the excitation energy of the system. When the residual
interaction between the particles is greater than the energy spacing between the
levels that it mixes, the eigenstates $|n\rangle$ become chaotic superpositions
of the basis states $|b\rangle$, constructed from the single-particle orbitals
(e.g., Slater determinants, for the Fermi system).
In this regime the coefficients in the eigenstate expansion,
\begin{align}\label{aa0}
|n\rangle = \sum _b C^{(n)}_b |b\rangle \,,
\end{align}
behave as uncorrelated random variables:
\begin{align}\label{sc0}
\overline{C^{(n)}_b}=0\,,\qquad \overline{C^{(m)}_a C^{(n)}_b}
=\delta_{mn}\delta_{ab}\, \overline{{C^{(n)}_b}^2}\,.
\end{align}
Note that we use indices $m$, $n$, etc., to denote the compound eigenstates,
and $a$, $b$, etc., for the basis states, and the averages are taken over
nearby eigenstates.

For the system under consideration the Hamiltonian matrix
$H_{ab}$ and the coefficients $C^{(n)}_b$ can be made real. We also assume
that the basis states and the eigenstates have the same
exact quantum numbers (e.g., the total angular momentum and parity, for a
spherically symmetric system), and the usual normalization condition applies:
$\sum_b |C_b^{(n)}|^2=\sum_n |C_b^{(n)}|^2=1$.

Besides \Eref{sc0}, the coefficients display a systematic dependence on the 
eigenstate energy. This dependence can be described by
(see, e.g., \cite{FGGK94}),
\begin{align}\label{aa2}
\overline{{C^{(n)}_b}^2} = \frac{D}{2\pi}\,
\frac{\Gspr}{(E_n-E_b)^2 +\Gspr^2/4}\,,
\end{align}
where $E_n$ is the energy eigenvalue of state $n$, $E_b\equiv H_{bb}$ is the
expectation energy of the basis state $b$, and $D$ is the mean energy spacing
between the eigenstates. The parameter
$\Gspr $ is the spreading width. It characterizes the size of the energy
interval in which the typical coefficients are close to maximum,
$C_b^{(n)}\sim 1/\sqrt{N}$, where $N=\pi \Gspr/(2D)$, is the 
number of {\em principal components}, i.e., the number of basis states that
contribute significantly to a given eigenstate. In the strong mixing regime,
$\Gspr \gg D$ and $N\gg 1$.
The spreading width can be calculated using the Golden Rule as
$\Gspr =2\pi \overline{|H_{ab}|^2}/D_b$, where $D_b$ is the mean spacing between
the states $b$ to which a given basis state $a$ is coupled
\cite{PhysRevA.86.022714}, or evaluated in relatively small scale
configuration-interaction calculations \cite{GS03}. Its values
in atomic systems range from $\sim 1$~eV in atoms, such as Ce \cite{FGGK94}, to
$\sim 10$~eV in multicharged ions, e.g., Au$^{24+}$ \cite{GGF99,GS03},
or W$^{q+}$ ($q=18$--24) \cite{PhysRevA.86.022714}.

\subsection{Coherent amplitudes}\label{subsec:coh}

Let us consider the process of photoexcitation of a many-electron atom or ion
$A$ from the ground state $|0\rangle$ to an excited state $|n\rangle$ above
the ionization limit. This leads to either autoionization
($A +\gamma \rightarrow A^+ + e$) or radiative quenching of the excited state
($A +\gamma \rightarrow A^* + \gamma'$). The corresponding amplitudes are
\begin{align}\label{VN1}
M^{\gamma e}_{k\eps }&=\sum_n \frac{\langle k,\eps |\hat V|n\rangle
\langle n|\hat D|0\rangle}{E_0+\omega-E_n +\frac{i}{2}\Gamma_n}\,,
\\ \label{VN2}
M^{\gamma \gamma'}_{m}&=\sum_n \frac{\langle m|\hat D|n\rangle
\langle n|\hat D|0\rangle}{E_0+\omega-E_n +\frac{i}{2}\Gamma_n}\,,
\end{align}
where  $\hat D$ is the electron-photon interaction operator, $\omega $ is the
photon energy, and $\hat V$ is the electron Coulomb interaction. The
first amplitude corresponds to the final state $|k\rangle$ of the ion $A^+$ and
an electron in the continuum state $|\eps \rangle$. The second amplitude
describes photon (Raman) scattering leading to the final atomic state $m$
and a photon $\gamma '$. The sums are over the compound eigenstates $n$ with
the energy $E_n$ and total width $\Gamma_n$ (due to both autoionization and
radiative decay).

Note that in considering the photon impact we neglect the possibility of
direct electron emission into the continuum. Such process will either produce
a distinct, smooth background for the resonant contributions, or, more likely
for complex targets, the continuum states will be strongly mixed with the
autoionizing resonances \cite{FGG96}. 

Using \Eref{aa0} in \Eref{VN1}, and averaging this amplitude over a
small energy interval containing many compound states $n$, gives the
\textit{coherent} part of the photoionization amplitude:
\begin{align}\label{photoMev}
\overline{M^{\gamma e}_{k\eps }} =\sum_{nd} \overline{{C^{(n)}_d}^2}
\frac{\langle k\eps |\hat V|d\rangle \langle d|\hat D|0\rangle}
{E_0+\omega-E_n+\frac{i}{2}\Gamma_n} \,,
\end{align}
where we also made use of \Eref{sc0}. The sum in \Eref{photoMev} is over the
compound states $n$ and basis states $d$. Since $\hat D$ is a one-body
operator, the matrix element $\langle d|\hat D|0\rangle $ is nonzero only for
the basis states $d$ in which one of the ground-state electrons is excited
by the photon (assuming that the ground state has a well-defined
configuration). Such states $d$ play the role of \textit{doorway} states for the
resonant photoabsorption process.

The mean spacing $D$ between the compound resonances is very small, which
allows one to replace summation over $n$ by integration,
\begin{align}\label{aa3}
\sum_n \rtw \int \frac{dE_n}{D}\,.
\end{align}
Using \Eref{aa2} in \Eref{photoMev}, we then obtain
\begin{equation}\label{aa4}
\overline{M^{\gamma e}_{k\eps }}=
\sum _d \frac{\langle k,\eps |\hat V|d\rangle\langle d|\hat D|0\rangle}
{E_0+\omega-E_d+\frac{i}{2}\Gspr }\,,
\end{equation}
where $\Gamma_n \ll \Gspr$ has been assumed. The latter relation is supported
by numerical calculations
\cite{GGF99,FGGH02,GS03,PhysRevA.86.022714,PhysRevA.88.062713},
which show that the natural width of compound states $\Gamma_n$ is several
orders of magnitude smaller than $\Gspr $.
Similarly, averaging the amplitude in \Eref{VN2} gives
\begin{align} \label{aa6}
\overline{M^{\gamma \gamma'}_{m}}=
\sum_d \frac{\langle m|\hat D|d\rangle \langle d|\hat D|0\rangle}
{E_0+\omega-E_d+\frac{i}{2}\Gspr}\,.
\end{align}

Equations \eqref{aa4} and \eqref{aa6} reveal the physical meaning of the
coherent amplitudes. They describe the excitation of the system into simple
doorway states $d$, which then decay directly into the final states.
(In the incoherent, ``stochastic'' contribution, the capture into a compound
state $n$ and its decay are due to different basis-state components $d$ and
$e$, see Sec. \ref{subsec:part}.) Doorway states are not the eigenstates of
the Hamiltonian, as they are mixed by the Coulomb interaction with other basis
states with two, three and more excited electrons. In the temporal picture of
the process, the photon initially excites one electron. This is followed by a
chain of electron interactions, until all the excitation energy is shared
between as many electrons as possible (cf. \fref{fig:ivr_rec}). This internal
decay of the doorway state on the time scale $\sim \hbar /\Gspr $ explains the
origin of the spreading width in the denominators of Eqs.~\eqref{aa4} and
\eqref{aa6}. The spreading width $\Gspr$ is similar to the quasiparticle width
in a solid where quasiparticles also decay into internal excitations of the
solid (see, e.g., Ref. \cite{PhysRevLett.78.2803} and references therein).

The doorway states for photoionization are single-electron excitations from the
ground state. The eigenstates of the Hamiltonian (i.e., the compound
resonances) contribute \textit{coherently} to each doorway state. Therefore,
this contribution is not included in the standard independent-resonance
approximation \cite{LL77,PBG92}.

\subsection{Total cross section}\label{subsec:tot}

The total cross section of the photon- or electron-induced reactions, averaged
over the compound resonances, can be found using the optical theorem
\cite{LL77}, from the elastic forward-scattering amplitude, e.g.,
for the photon-induced case,
$\st^\gamma \propto {\rm Im}\, M^{\gamma \gamma }_0$. Averaging this
relation over the compound resonances involves the
coherent contribution (\ref{aa6}) for $|m\rangle=|0\rangle$, and we have
\begin{align}\label{sigmatcoh}
\st^\gamma \propto {\rm Im}\,\overline{M^{\gamma \gamma}_0}=
\frac{1}{2}\sum_d \frac{|\langle d|\hat D|0\rangle|^2\Gspr }
{(E_0+\omega-E_d)^2+\Gspr ^2/4}\,,
\end{align}
where the sum is over the doorway states $d$.

Note that the integral contribution of each of the doorway states in
\Eref{sigmatcoh} ($\int \st d\omega $) is independent of $\Gspr $. The total
photoabsorption cross
section is given by the sum of the single-particle (i.e., doorway)
contributions. The only manifestation of the strong mixing and chaotic dynamics
in the system is the broadening of these single-particle peaks by
$\Gspr $ (which is much greater than the natural widths of the single-particle
excitations).

A familiar example of this picture is the infrared absorption
spectra of molecules, which are dominated by characteristic peaks of various
modes. A low-resolution measurement of the total cross section will not
reveal any features related to the strong mixing or IVR, which take place
after the absorption of the photon.

As a  consistency check we can obtain the result of \Eref{sigmatcoh} starting
from the sum over compound states in \Eref{VN2}. Setting $m=0$, we have
\begin{align}\label{sigmat}
{\rm Im}\,M^{\gamma \gamma}_0
=\frac{1}{2}\sum _n \frac{|\langle n|\hat D|0\rangle|^2 \Gamma_n }
{(E_0+\omega-E_n)^2+\Gamma_n ^2/4}\,.
\end{align}
Using Eqs. \eqref{aa0} and \eqref{sc0}, one obtains
\begin{align}\label{eq:ImMgg}
{\rm Im}\,\overline{M^{\gamma \gamma}_0}
=\frac{1}{2}\sum_{nd} \overline{{C^{(n)}_d}^2}
\frac{|\langle d|\hat D|0\rangle|^2 \Gamma_n}
{(E_0+\omega-E_n)^2+\Gamma_n ^2/4}\,,
\end{align}
and applying Eqs.~\eqref{aa2} and \eqref{aa3} again leads to \Eref{sigmatcoh}.
One can also obtain \Eref{sigmatcoh} by averaging
\Eref{sigmat} over a photon energy interval $\Delta \omega$,
$\Gamma_n\ll \Delta \omega\ll \Gspr $, containing a large number of resonances
$\Delta \omega/D$ (i.e., integrating the resonant contributions over
$\omega$ instead of $E_n$).

A calculation similar to that in Sec. \ref{subsec:coh}, yields coherent
amplitudes of the electron-induced processes, i.e., photorecombination
($A^{q+} + e\rightarrow A^{(q-1)+} + \gamma $) and electron scattering
($A^{q+} + e\rightarrow {A^{q+}}^* + e'$) via compound resonances:
\begin{align}\label{Megamma}
\overline{M^{e \gamma}_{im}} &= \sum_d
\frac{\langle m|\hat D|d\rangle\langle d|\hat V|i,\eps \rangle}
{E_i+\eps -E_d+\frac{i}{2}\Gspr }\,,\\
\label{Mee}
\overline{M^{e e'}_{ik}} &= \sum_d
\frac{\langle k, \eps '|\hat V|d\rangle\langle d|\hat V|i,\eps \rangle}
{E_i+\eps -E_d+\frac{i}{2}\Gspr }\,.
\end{align}
Here the doorway states $d$ are dielectronic excitations of the ion
$A^{(q-1)+}$, produced by capturing the incident electron simultaneously with
excitation of an electron of the target $A^{q+}$. In Eqs.~(\ref{Megamma})
and (\ref{Mee}), $i$ is the initial (e.g., ground) state of the target ion,
$m$ is the final state of the ion $A^{(q-1)+}$, and $k$ is the final state
of ${A^{q+}}^*$.

In analogy to \Eref{sigmatcoh}, the averaged total resonant electron-impact
cross section is
\begin{align}\label{sigmatecoh}
\st^e \propto {\rm Im}\,\overline{M^{ee}_{ii}}=\frac{1}{2}\sum_d
\frac{|\langle d|\hat V|i,\eps \rangle|^2 \Gspr }
{(E_i+\eps -E_d)^2+\Gspr ^2/4}\,.
\end{align}
It describes all processes following the capture of an electron in the
dielectronic doorway states, broadened (via $\Gspr$) by
multiconfigurational mixing which defines the compound eigenstates.

Equations \eqref{sigmatcoh} and \eqref{sigmatecoh} can be written in the
familiar Breit-Wigner form by replacing the squared matrix elements by the
corresponding partial widths for the decay of the doorway state. Hence, we
introduce the radiative width
$\Gamma_{d\rightarrow 0}^{(r)} \propto |\langle d|\hat D|0\rangle|^2$, and
the autoionization width
$\Gamma_{d\rightarrow i}^{(a)} \propto  |\langle d|\hat V|i,\eps \rangle|^2$.
It is also natural to add the total radiative width $\Gamma_{d}^{(r)}$ and total
autoionization width $\Gamma_{d}^{(a)}$ of the doorway to its spreading width,
to account for all decay modes of this state. The total width of the doorway
state then is $\Gamma_{d}=\Gspr + \Gamma_{d}^{(r)}+\Gamma_{d}^{(a)}$, and
the cross sections are given by
\begin{align}\label{sigmaradtotal}
\st^\gamma &\propto \sum_d \frac{\Gamma_{d\rightarrow 0}^{(r)} \Gamma_d}
{(E_0+\omega-E_d)^2+\Gamma_d^2/4},\\
\label{sigmaautototal}
\st^e &\propto \sum_d \frac{ \Gamma_{d\rightarrow i}^{(a)} \Gamma_d}
{(E_0+\omega-E_d)^2+\Gamma_d^2/4}.
\end{align}
In this form it is easy to restore the correct prefactor in these equations,
by comparison with the standard Breit-Wigner formula \cite{LL77}.

In Sec. \ref{subsec:coh} and above, the doorways states were introduced as
particular types of basis states selected by the process under consideration.
To make Eqs. \eqref{sigmatcoh} and \eqref{sigmatecoh} [or (\ref{sigmaradtotal})
and (\ref{sigmaautototal})] more accurate for application to real systems, one
can diagonalize the Hamiltonian matrix in the subspace of the doorway states.
This should supply more accurate energies $E_d$ and amplitudes involving
the doorways. In complex systems the doorways are only a small part of the
total Hilbert space in the energy range of interest, making this task feasible.

Note that Eqs. (\ref{sigmaradtotal}) and (\ref{sigmaautototal}) provide
interpolation formulas for the total cross sections. They can describe
a transition from the chaotic compound resonance regime, in which
$\Gamma_d \approx \Gspr$, to the simple resonance regime
$\Gamma_d\approx \Gamma_{d}^{(r)}+\Gamma_{d}^{(a)}$ (in which the ``doorway''
states do not spread).
For $\Gamma_{d}^{(a)}+\Gamma_{d}^{(r)} \gg \Gspr$, the doorway state has no time
to excite other electrons and is decoupled from the compound resonances. This
can also be explained using perturbation theory. In this case the energy
difference between a doorway state $d$ and a compound state $n$,
$E_d -E_n - i (\Gamma_{d}^{(a)}+\Gamma_{d}^{(r)})/2$, is dominated by the
imaginary part and becomes larger than the the matrix element of the residual
interaction $\hat V$, which can mix $d$ and $n$, i.e.,
$\langle n|\hat V |d\rangle /\Gamma_{d}^{(a)}\ll 1$. (Except for the very highly
charged ions, $\Gamma_{d}^{(a)}\gg \Gamma_{d}^{(r)}$ for the dielectronic
states.)

Numerical calculations for W$^{19+}$ and Au$^{24+}$ show that
$\Gamma_d^{(a)} \ll \Gspr $, and the electron recombination processes in such
ions are dominated by the many-electron compound resonances
\cite{FGGH02,PhysRevA.86.022714,PhysRevA.88.062713} (see below).

\subsection{Partial cross sections}\label{subsec:part}
 
The total width of a resonance $n$ is the sum of its partial widths over all
final states or decay channels, $\Gamma _n=\sum _f\Gamma _n^{(f)}$. In the
independent-resonance approximation the partial cross section $\sigma _f$ for
channel $f$ (averaged over the resonances) can be obtained by multiplying
the total cross section $\st$ by the average ratio of the corresponding partial
width $\Gamma_n^{(f)}$ to the total width $\Gamma_n$. In most cases the
compound state $n$ can decay into many final states, which suppresses the
fluctuations of $\Gamma_n$ \cite{FGGH02,PhysRevA.86.022714,PhysRevA.88.062713},
and one obtains
\begin{align}\label{sigmapartial}
\sigma _f \approx \st \overline{\Gamma_n^{(f)}}\big/ \overline{\Gamma_n}\,.
\end{align}
However, in this approximation one misses a specific, coherent contribution to
the partial cross section, which is calculated below. 

The resonance-averaged cross section (or probability) of a process is
proportional to the modulus squared amplitude, $P=\overline{|M|^2}$. When
analyzing this quantity, it is convenient
to separate out the coherent term, $P_{\rm coh}=|\overline{M}|^2$. The
remaining part then represents the \textit{stochastic} contribution:
$P_{\rm sto} = \overline{|M|^2}-|\overline{M}|^2$.

Let us consider photoionization as an example. The corresponding
resonance-averaged probability $P^{\gamma e}$ is found by taking the squared
modulus of the amplitude $M^{\gamma e}_{k\eps }$ from \Eref{VN1}:
\begin{widetext}
\begin{align}\label{eq:M2}
|M^{\gamma e}_{k\eps }|^2=\sum_{n,n'}
\frac{\langle 0|\hat D|n'\rangle \langle n' |\hat V|k,\eps \rangle
}{E_0+\omega-E_{n'} -\frac{i}{2}\Gamma_{n'}}\,
\frac{\langle k,\eps |\hat V|n\rangle
\langle n|\hat D|0\rangle}{E_0+\omega-E_n +\frac{i}{2}\Gamma_n}\,,
\end{align}
Each of the four matrix elements in this expression involves one compound state
($n$ or $n'$), which can be expanded as in \Eref{aa0}. After this, averaging of
\Eref{eq:M2} reduces to finding the averaged product of four expansion
coefficients:
\begin{align}\label{sc1}
\overline{C^{(n')}_a C^{(n')}_b C^{(n)}_c C^{(n)}_d}
=\delta_{ab}\overline{{C^{(n')}_b}^2}\,\delta_{cd}\overline{{C^{(n)}_d}^2} 
+\delta_{n'n}\delta_{ad}\overline{{C^{(n)}_d}^2}\,
\delta_{bc}\overline{{C^{(n)}_b}^2}
+\delta_{n'n}\delta_{ac}\overline{{C^{(n)}_a}^2}\,
\delta_{bd}\overline{{C^{(n)}_d}^2}
\,,
\end{align}
which follows from \Eref{sc0}. Hence, the average of \Eref{eq:M2} is the sum
of three distinct terms:
\begin{align}\label{sc2}
P^{\gamma e}=\left|\sum_{nd}\overline{{C^{(n)}_d}^2}
\frac{\langle k,\eps|\hat V|d\rangle\langle d|\hat D|0\rangle}
{E-E_n+\frac{i}{2}\Gamma_n}\right|^2
+\sum_{nbd}\overline{{C^{(n)}_b}^2}\,\overline{{C^{(n)}_d}^2}
\frac{|\langle k,\eps |\hat V|b \rangle |^2 |\langle d|\hat D|0\rangle |^2}
{(E-E_n)^2+\Gamma_n^2/4}
+\sum_{n}\left|\sum_d \overline{{C^{(n)}_d}^2}
\frac{\langle d|\hat V|k,\eps \rangle \langle d|\hat D|0\rangle}
{E-E_n-\frac{i}{2}\Gamma_n}\right|^2 \,,
\end{align}
where $E=E_0+\omega $ is the total energy of the system.
\end{widetext}

The first term on the right hand side of \Eref{sc2} is the coherent
contribution,
cf. \Eref{photoMev}. The second term corresponds to the independent resonance
approximation, and is usually the only term considered \cite{LL77,PBG92}.
The weights given by the mean-squared coefficients, which multiply the modulus
squared matrix elements for autoionization and photoabsorption,
link the corresponding partial widths of the compound and doorway states:
\begin{align}\label{eq:Gamrn}
\Gamma_{n\rightarrow 0}^{(r)}&=
\sum _d \overline{{C^{(n)}_d}^2}\Gamma_{d\rightarrow 0}^{(r)},\\ \label{eq:Gaman}
\Gamma_{n\rightarrow k}^{(a)} &=
\sum _b \overline{{C^{(n)}_b}^2}\Gamma_{b\rightarrow k}^{(a)}
\end{align}
The last term in \Eref{sc2} is the remaining part of the stochastic
contribution, and we call it the residual stochastic term. The stochastic
contribution thus consists of the independent resonance (IR) contribution and
the residual stochastic term, $P_{\rm sto} =P_{\rm IR}+P_{\rm res}$.

Using Eqs.~\eqref{aa2} and \eqref{aa3} [or averaging  \Eref{sc2}  over  the
energy interval $\Delta \omega$, as explained below \Eref{eq:ImMgg}], we find
the coherent contribution to the partial cross section,
\begin{align}\label{Pcoh}
P^{\gamma e}_{\rm coh} = \left|\sum_d
\frac{\langle k,\eps |\hat V|d\rangle \langle d|\hat D|0\rangle}
{E_0+\omega-E_d+\frac{i}{2}\Gspr } \right|^2\,,
\end{align}
the independent-resonance contribution,
\begin{align}
P^{\gamma e}_{\rm IR} =
\frac{D}{2 \pi \Gamma_n}&\sum_b
\frac{|\langle k,\eps |\hat V|b\rangle|^2\Gspr}
{(E_0+\omega-E_b)^2+\Gspr^2/4}\nonumber \\ \label{PIR}
\times &\sum _d \frac{|\langle d|\hat D|0\rangle|^2\Gspr}
{(E_0+\omega-E_d)^2+\Gspr^2/4}\,,
\end{align}
and the residual stochastic contribution,
\begin{align}\label{Pres}
P^{\gamma e}_{\rm res} = \frac{D}{2 \pi \Gamma_n}\left|\sum_{d}
\frac{\langle d|\hat V|k,\eps \rangle \langle d|\hat D|0\rangle \Gspr }
{(E_0+\omega-E_d)^2+\Gspr^2/4} \right|^2 \,.
\end{align}
In these expressions the matrix elements and sums involve only doorway states.
Compound resonances have been eliminated from the sums, and only affect the
result through the parameters such as $\Gspr$, the mean level spacing $D$ and
the compound state width $\Gamma_n$. Similar
expressions can be obtained for the probabilities of the photon and electron
scattering and electron recombination. These formulae are suitable for the
numerical calculations of the resonance-averaged cross sections. Conversion of
these equations to the cross sections involves kinematic factors, whose precise
form depends on the normalization of the electron continuum states
$\varepsilon $ and electromagnetic transition operator $\hat D$. 

\section{Comparison of the coherent and stochastic contributions}

\subsection{Photoionization}

Let us compare the magnitudes of the three contributions to the
resonance-averaged probability of photoionization
($A +\gamma \rightarrow A^+ +e$), Eqs.~(\ref{Pcoh})--(\ref{Pres}).
There are two reasons for the possible suppression of the coherent and residual
contributions in comparison with the independent-resonance term.

The first point to note is that the basis
(doorway) states which contribute to the sums over $b$ and $d$ in the
independent-resonance contribution $P^{\gamma e}_{\rm IR}$, \Eref{PIR}, are
in general quite different. The operator $\hat D$, which couples the ground
state $|0\rangle $ with state $|d\rangle $ is a one-body operator. Hence, the
photoabsorption doorway states $d$ are single-electron excitations from the
ground state. On the other hand, the two-body Coulomb interaction which
couples the final state $|k,\eps \rangle $ with $|b\rangle $, favours
dielectronic (doubly-excited) doorway states $b$. The level density of such
states is much higher than that of the single-electron excitations. This means
that the number of terms which contribute effectively to the sum over $b$,
$N_b\sim \Gspr /D_b$, is much greater than the number of terms which contribute
to the sum over $d$, $N_d\sim \Gspr /D_d$ (where $D_b$ and $D_d$ are the mean
spacing between the corresponding doorway states, $D_b\ll D_d$). We thus see
that the sum in \Eref{PIR} contains $\sim N_bN_d$ positive terms.

In contrast, in both the coherent and residual stochastic parts
$P^{\gamma e}_{\rm coh}$ and $P^{\gamma e}_{\rm res}$, Eqs.~(\ref{Pcoh}) and
(\ref{Pres}), the same doorway $d$ appears in both matrix
elements. As a result, these sums contain $\sim N_d^2$ terms. Besides this,
only $\sim N_d$ of these terms (i.e., the diagonal ones) are definitely
positive, while the remaining interference terms can have different signs.
The expressions for the independent-resonance and the residual stochastic
contributions, Eqs.~(\ref{PIR}) and (\ref{Pres}), contain the same prefactors,
and we see that the residual contribution is suppressed as
$P^{\gamma e}_{\rm res}/P^{\gamma e}_{\rm IR}\sim 1/N_b$.

The situation with the coherent contribution is not so simple. According to the
above estimates, we have
\begin{align}\label{eq:cohIR}
\frac{P^{\gamma e}_{\rm coh}}{P^{\gamma e}_{\rm IR}}\sim
\frac{1}{N_b}\,\frac{\Gamma_n}{D}\sim \frac{D_b \Gamma_n}{\Gspr D}.
\end{align}
If the compound resonances have a small number of decay channels, then
$\Gamma_n\ll D$ would normally hold (see Appendix B of Ref.~\cite{FGG96} and
references therein), and the independent-resonance contribution dominates.
However, for the compound states which lie above the ionization threshold, the
number of decay channels can be large. In this case one can have
$\Gamma _n\sim D$ \cite{FGGH02}, or even observe strongly overlapping
resonances with $\Gamma _n\gg D$. This means that there could be cases in which
the coherent contribution is important.

To illustrate the role of doorways, the photoionization cross section of
Xe$^{q+}$ ions ($q=4$--6) in the energy range $\omega =90$--100~eV is
dominated by a prominent narrow maximum due to the $4d\rightarrow 4f$
transition \cite{AGG06}.
Xe$^{6+}$ is a closed-shell system, and the $4d-4f$ peak in this system
appears as a structureless single-particle peak. In the open-shell Xe$^{5+}$ and
Xe$^{4+}$, the $4d-4f$ peak becomes progressively more fragmented, due to
mixing between the $4d-4f$ doorway and other electronic excitations.

\subsection{Photon scattering}

For photon scattering ($A +\gamma \rightarrow A^* +\gamma'$), the
coherent, independent-resonance, and residual stochastic contributions
are obtained by averaging $|M_m^{\gamma \gamma '}|^2$, where
$M_m^{\gamma \gamma '}$ is given by \Eref{VN2}. The result is given by
expressions similar to those in Eqs.~(\ref{Pcoh})--(\ref{Pres}):
\begin{align}\label{Pcohgg}
P^{\gamma \gamma'}_{\rm coh} &= \left|\sum_d
\frac{\langle m|\hat D|d\rangle  \langle d|\hat D|0\rangle}
{E_0+\omega-E_d+\frac{i}{2}\Gspr } \right|^2\,,\\
P^{\gamma \gamma'}_{\rm IR} &=\frac{D}{2 \pi \Gamma_n}\sum_b
\frac{|\langle m|\hat D|b\rangle |^2\Gspr}
{(E_0+\omega-E_b)^2+\Gspr^2/4}\nonumber \\ \label{PIRgg}
&\times \sum _d \frac{|\langle d|\hat D|0\rangle|^2\Gspr}
{(E_0+\omega-E_d)^2+\Gspr^2/4}\,, \\ \label{Presgg}
P^{\gamma \gamma'}_{\rm res} &= \frac{D}{2 \pi \Gamma_n}\left|\sum_{d}
\frac{\langle d|\hat D|m\rangle  \langle d|\hat D|0\rangle \Gspr }
{(E_0+\omega-E_d)^2+\Gspr^2/4} \right|^2 \,.
\end{align}
For elastic ($m=0$)
or weakly inelastic scattering (e.g., when the final state $m$ belongs to the
same electronic configuration as the initial state 0), same doorways
$d$ will be available in the sums for $P^{\gamma \gamma'}_{\rm coh}$
and $P^{\gamma \gamma'}_{\rm res}$, so that the latter is suppressed as
$1/N_d$ relative to $P^{\gamma \gamma'}_{\rm IR}$. Simple single-electron
excitation doorways do not have a dense spectrum, which means that $N_d$ may be
small, making all three contributions comparable.

On the other hand, if the energy of the incident photon is sufficiently large,
inelastic (Raman) photon scattering becomes much more prominent due to the
availability of many excited final states $m$. The majority of them will share
no or few doorways with the initial state 0, which means that both
$P^{\gamma \gamma'}_{\rm coh}$ and $P^{\gamma \gamma'}_{\rm res}$ will be strongly
suppressed in comparison with $P^{\gamma \gamma'}_{\rm IR}$. The same conclusion
is true if we consider the total photon scattering cross section summed over
the final states $m$.

To make the comparison clearer, we can present our results in a conventional
Breit-Wigner form by replacing the squared matrix elements by the
corresponding partial widths. For example, the coherent contribution
\eqref{Pcohgg} to the total photon scattering cross section is
\begin{align}\label{Pgg_coh}
P^{\gamma \gamma'}_{\rm coh}&= \sum_{dm}
\frac{|\langle m|\hat D|d\rangle|^2\,|\langle d|\hat D|0\rangle|^2}
{(E_0+\omega-E_d)^2+\Gspr^2/4} \\ \label{aa11}
&\propto \sum_{d}\frac{\Gamma_{d\rightarrow 0}^{(r)}\Gamma_d^{(r)}}
{(E_0+\omega-E_d)^2+\Gspr^2/4}\,,
\end{align}
where the total radiative width of the doorway state $d$ is
\begin{align}
\label{aa13}
\Gamma_d^{(r)}&\propto \sum _m |\langle d|\hat D|m\rangle|^2
\approx \sum _e |\langle d|\hat D|e\rangle|^2\,.
\end{align}
In \Eref{Pgg_coh} we neglected the interference terms between different doorway
states in (\ref{Pcohgg}), since their contribution is strongly suppressed
($\sim N_m^{-1/2}$) after summation over the large number $N_m$ of compound
states $m$ populated after the emission of the final-state photon. Note also
that we have replaced the sum over the compound states in the total radiative
width \eqref{aa13} by the sum over the basis states $e$, owing to
normalization $\sum_m |C_e^{(m)}|^2=1$. As a result, the final expression
in \Eref{aa11} includes only the matrix elements between doorway states and
relatively simple states $|0\rangle$ and  $|e\rangle$.

Equation \eqref{aa11} describes the Breit-Wigner-like contributions of the
doorway states $d$ to the coherent part of the photon scattering cross section.
Comparing with the total cross section (\ref{sigmaradtotal}), we see that the
coherent contribution is suppressed by the ratio
$\Gamma_d^{(r)}/\Gamma_d\approx \Gamma_d^{(r)}/\Gspr \ll 1$.
In a similar way, the independent-resonance contribution can be written as
\begin{align}\label{eq:PIRgg}
P_{\rm IR}^{\gamma \gamma '}\propto \frac{\Gamma _n^{(r)}}{\Gamma _n}
\sum _d\frac{\Gamma_{d\rightarrow 0}^{(r)}\Gspr }
{(E_0+\omega-E_d)^2+\Gspr^2/4},
\end{align}
where
\begin{align}\label{eq:Gamrn1}
\Gamma _n^{(r)}=\sum _m\Gamma _{n\rightarrow m}^{(r)}=
\sum _{mb}\overline{{C^{(n)}_b}^2}\Gamma_{b\rightarrow m}^{(r)},
\end{align}
[cf. Eqs. (\ref{PIR}) and (\ref{eq:Gamrn})]. Compared to the total cross 
section (in which $\Gamma _d\approx \Gspr $), the independent-resonance
contribution \Eref{eq:PIRgg} contains an extra factor
$\Gamma _n^{(r)}/\Gamma _n$, which is the branching ratio for the radiative
decay of the resonances. Since $\Gamma _n^{(r)}\sim \Gamma_d^{(r)}$ (for the
doorways represented in $n$), the ratio
$P^{\gamma \gamma'}_{\rm coh}/P_{\rm IR}^{\gamma \gamma '}\sim\Gamma _n/\Gspr \ll 1$,
i.e., the coherent contribution is suppressed in comparison with the
the independent-resonance term.

\subsection{Electron scattering}

Considering electron scattering ($A +e \rightarrow A^* +e'$), the three
contributions to the resonance-averaged cross section are
\begin{align}\label{Pcohee}
P^{ee'}_{\rm coh} &= \left|\sum_d
\frac{\langle k, \eps '|\hat V|d\rangle  \langle d|\hat V|i,\eps \rangle}
{E_i+\eps -E_d+\frac{i}{2}\Gspr } \right|^2\,,\\
P^{ee'}_{\rm IR} &=\frac{D}{2 \pi \Gamma_n}\sum_b
\frac{|\langle k, \eps '|\hat V|b\rangle  |^2\Gspr}
{(E_i+\eps -E_b)^2+\Gspr^2/4}\nonumber \\ \label{PIRee}
&\times \sum _d \frac{|\langle d|\hat V|i,\eps \rangle|^2\Gspr}
{(E_i+\eps -E_d)^2+\Gspr^2/4}\,, \\ \label{Presee}
P^{ee'}_{\rm res} &= \frac{D}{2 \pi \Gamma_n}\left|\sum_{d}
\frac{\langle d|\hat V|k, \eps '\rangle \langle d|\hat V|i,\eps \rangle\Gspr }
{(E_i+\eps -E_d)^2+\Gspr^2/4} \right|^2 \,.
\end{align}
For low incident electron energies, one can only have elastic or quasielastic
scattering, when state $k$ is identical or similar to $i$. In this case the
suppression of the coherent and residual contributions is $\sim 1/N_d$.
However, the doorways involved in electron capture and re-emission are
dielectronic excitations of the compound atom or ion. Their level density is
higher than that of single-electron excitations, leading to greater values of
$N_d$ and stronger suppression than in photon scattering. At higher incident
electron energies more final states $k$ become available. Such states will have
fewer common doorways with the initial states and the relative importance of
the $P_{\rm IR}^{ee'}$ will increase further.

\subsection{Electron-ion recombination}

Last but not least, electron recombination
($A^{q+} +e \rightarrow A^{(q-1)+} +\gamma$) is quite special. For complex
targets, many final states (channels) are available even at the lowest
incident electron energy \footnote{The number of finite states is infinite
even for simple target ions, as they include Rydberg states for the
incident electron. However, this is a feature of the single-electron dynamics,
and it does not lead to enhancements related to many-electron excitations
produced in complex systems.}. To obtain the total recombination
cross section, one needs to sum over all final states $m$ of the ion
$A^{(q-1)+}$. The three contributions to the reaction probability then are
\begin{align}\label{Pcoheg}
P^{e\gamma}_{\rm coh} &= \sum _m\left|\sum_d
\frac{\langle m|\hat D|d\rangle  \langle d|\hat V|i,\eps \rangle}
{E_i+\eps -E_d+\frac{i}{2}\Gspr } \right|^2\,,\\
P^{e\gamma}_{\rm IR} &=\frac{D}{2 \pi \Gamma_n}\sum _m \sum_b
\frac{|\langle m|\hat D|b\rangle  |^2\Gspr}
{(E_i+\eps -E_b)^2+\Gspr^2/4}\nonumber \\ \label{PIReg}
&\times \sum _d \frac{|\langle d|\hat V|i,\eps \rangle|^2\Gspr}
{(E_i+\eps -E_d)^2+\Gspr^2/4}\,, \\ \label{Preseg}
P^{e\gamma}_{\rm res} &= \frac{D}{2 \pi \Gamma_n}\sum _m\left|\sum_{d}
\frac{\langle d|\hat D|m\rangle \langle d|\hat V|i,\eps \rangle\Gspr }
{(E_i+\eps -E_d)^2+\Gspr^2/4} \right|^2 \,.
\end{align}
For the majority of states $m$, the doorways $b$ and $d$ in the
two matrix elements in Eq.~(\ref{PIReg}) will be different. This means that both
the coherent and the residual contributions, in which $b=d$, are strongly
suppressed relative to the independent-resonance contribution.

Similarly to \Eref{eq:PIRgg}, the recombination probability
$P^{e\gamma }_{\rm IR}$ can be written in terms of the widths,
\begin{align} \label{eq:PIReg1}
P^{e\gamma }_{\rm IR} \propto \frac{\Gamma _n^{(r)}}{\Gamma _n}
\sum_d \frac{ \Gamma_{d\rightarrow i}^{(a)} \Gspr }{(E_i+\eps-E_d)^2+\Gspr^2/4}.
\end{align}
This large ratio of the number of open channels for photon- and
electron-emission reactions at low incident electron energies also explains
the enhancement of the fluorescent yield in the resonant electron capture up
to nearly 100\% (i.e., $\Gamma_n^{(r)}/\Gamma_n \approx 1$)
\cite{GGF99,FGGH02,PhysRevA.86.022714,PhysRevA.88.062713}. In this
case electron recombination dominates in the total cross section
\Eref{sigmaautototal} of the electron collisions with highly charged ions, and
the total cross section can be calculated using the IR stochastic contribution
only. On the other hand, the total cross section is expressed via the imaginary
part of the coherent elastic amplitude [cf. \Eref{sigmatecoh}]. Thus, the
optical theorem establishes a relation between the coherent and stochastic
contributions.

\begin{figure}[ht!]
\includegraphics*[width=8.5cm]{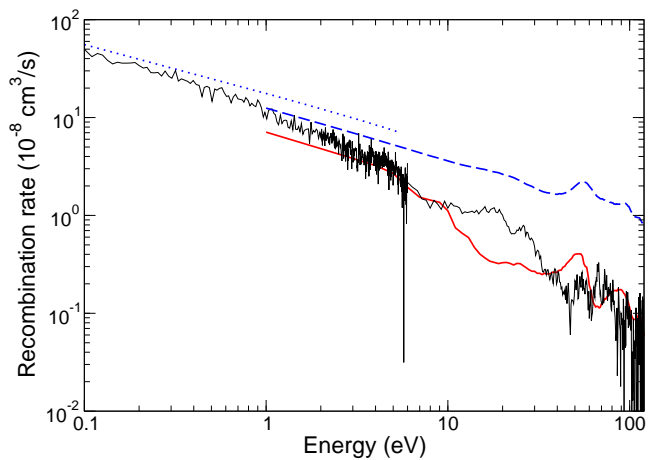}
\caption{(Color online) The graph shows the total rate of electron resonant
capture by W$^{20+}$, calculated using statistical theory (dotted blue line,
Ref.~\cite{PhysRevA.86.022714}; dashed blue line,
Ref.~\cite{PhysRevA.88.062713}), and the calculated recombination rate (thick
solid red line), obtained by including the effect of the nonunit fluorescence
yield \cite{PhysRevA.88.062713}. Thin solid black line is the experimental
recombination rate from Ref.~\cite{PhysRevA.83.012711}.}
\label{fig:W19+}
\end{figure}

To illustrate the effect of the fluorescence yield, \fref{fig:W19+} shows
the calculated total resonant capture rate and the photorecombination rate for
electron collisions with $W^{20+}$. Working equations for the process of
electron-ion recombination, in which the quantities of interest are expressed
in terms of two-particle radial Coulomb integrals, angular momentum algebra
coefficients and sums over single-particle states, can be found in Refs.
\cite{FGGH02,PhysRevA.86.022714,PhysRevA.88.062713}. The small difference
between the resonant
capture rate from Ref. \cite{PhysRevA.86.022714} (calculated for $\eps =1$~eV
and plotted assuming $1/\eps $ energy dependence of the cross section)
and \cite{PhysRevA.88.062713} is due to a slightly different numerical
procedure. Suppression of the recombination rate with respect to the total
resonant capture rate, due to the factor $\Gamma_n^{(r)}/\Gamma_n $, is clearly
visible (see Refs. \cite{PhysRevA.86.022714,PhysRevA.88.062713} for the working
equations used). The difference in the detailed energy dependence of the
calculated and measured recombination rate is likely due to inaccurate energies
of the dielectronic doorways in the calculation.

It is interesting that the same mechanism that leads to strongly enhanced
recombination in $A^{q+} +e$ collisions, should strongly suppress
photoionization of $A^{(q-1)+}$ at photon energies close to threshold.
Re-writing the corresponding IR contribution \Eref{PIR} in terms of widths,
\begin{align} \label{eq:PIRge}
P_{\rm IR}^{\gamma e}\propto \frac{\Gamma _n^{(a)}}{\Gamma _n}
\sum _d\frac{\Gamma_{d\rightarrow 0}^{(r)}\Gspr }
{(E_0+\omega-E_d)^2+\Gspr^2/4},
\end{align}
where $\Gamma _n^{(a)}$ is the total autoionization width of state $n$, we
see that in ions such as Au$^{24+}$ or W$^{19+}$, in which
$\Gamma _n^{(r)}>\Gamma_n^{(a)}$ near threshold, Raman scattering,
\Eref{eq:PIRgg}, will be favoured over ionization. A possible way of observing
this effect in experiment is to measure and compare the total photoabsorption
and photoionization cross sections. Alternatively, one can measure the spectrum
of secondary photon from the resonant Raman scattering, varying the primary
photon energy across the ionization threshold.

\section{Conclusions}

In this work we have investigated the role of doorway states in electron- and
photon-induced reactions mediated by strongly mixed compound resonances.
Our analysis shows that the resonance-averaged total reaction cross sections
$\st$ are given by the coherent contributions of the compound resonances.
These cross sections are expressed in terms of doorway resonances
(i.e., simple states coupled directly to the initial state of the target).
The only difference with the standard approach for dielectronic recombination
and photoionization is that the doorway resonances are broadened by the
spreading width $\Gspr$, which described their coupling to the dense spectrum
of chaotic compound states.

The situation with the partial cross sections is more complicated. For
processes such as radiative electron capture (photorecombination) or photon
scattering at energies which place the system in the strong-mixing regime,
the number of the decay channels is very large. As a result, the stochastic
contribution, corresponding to the independent-resonance approximation
dominates. The (resonance-averaged) partial cross section can then be
calculated from the total cross section by including the appropriate
branching ratio, e.g.,
$\sigma ^e_r=\Big( \overline{\Gamma_n^{(r)}}\big/\overline{\Gamma_n}\Big)\st ^e$,
for photorecombination. Here the ratio of the radiative and total widths of
the compound resonances and the electron capture cross section $\st^e$ can be
calculated as in Ref.~\cite{PhysRevA.88.062713}. On the other hand, if the
process leads to electron emission (photoionization, electron scattering) and
the number of decay channels is small (i.e., the energy does not exceed
threshold by much), the independent-resonance approximation may be deficient
and the coherent contribution may need to be included.

\acknowledgments
This work was supported by the  Australian Research Council and Russian
Foundation for Basic Research Grant No.\ 14-02-00241. We thank J. Berengut,
V. Dzuba and C. Harabati for useful discussions.



%

\end{document}